\documentclass[preprint,superscriptaddress,showpacs,preprintnumbers,amsmath,amssymb]{revtex4-1}

\usepackage{amsmath}
\usepackage{amssymb}
\usepackage{graphicx}
\usepackage{morefloats}
\usepackage[usenames,dvipsnames]{xcolor}
\usepackage{blkarray}
\usepackage{verbatim}
\usepackage{hyperref}
\usepackage{color}

\begin{document}

\title{The Host-Pathogen Game:\\ an evolutionary approach to biological competitions}

\author{Marco Alberto Javarone}
\email{marcojavarone@gmail.com}

\affiliation{School of Computing, University of Kent, Chatham Maritime, UK}
\affiliation{nChain Ltd, London W1W 8AP, UK}

\date{\today}

\begin{abstract}
We introduce a model called Host-Pathogen game for studying biological competitions. Notably, we focus on the invasive dynamics of external agents, like bacteria, within a host organism.
The former are mapped to a population of defectors that aim to spread in the extracellular medium of the host. In turn, the latter is composed of cells, mapped to a population of cooperators, that aim to kill pathogens.
The cooperative behavior of cells is fundamental for the emergence of the living functions of the whole organism, since each one contributes to a specific set of tasks. So, broadly speaking, their contribution can be viewed as a form of energy.
When bacteria are spatially close to a cell, the latter can use a fraction of its energy to remove them.
On the other hand, when bacteria survive an attack, they absorb the received energy, becoming stronger and more resistant to further attacks. In addition, since bacteria play as defectors, their unique target is to increase their wealth, without supporting their own kind.
As in many living organisms, the host temperature plays a relevant role in the host-pathogen equilibrium. For instance, in animals like human beings, a neural mechanism triggers the increasing of the body temperature in order to activate the immune system. Here, cooperators succeed once bacteria are completely removed while, in the opposite scenario, the host undergoes a deep invasive process, like a blood poisoning.
Results of numerical simulations show that the dynamics of the proposed model allow to reach a variety of states. At a very high level of abstraction, some of these states seem to be similar to those that can be observed in some living systems.
Therefore, to conclude, we deem that our model might be exploited for studying further biological phenomena.
\end{abstract}

\maketitle
\section{Introduction}
In the last years, scientists belonging to different communities investigated socio-economic systems and biological phenomena under the lens of evolutionary game theory (hereinafter EGT)~\cite{nowak01,perc01,perc02,szolnoki01,masuda01,tomassini01,moreno01,traulsen01,szabo01,friedman01,nowak03,shuster01,frey01,javarone01,szolnoki02,javarone04}.
In general, studying the evolution of a population and identifying strategies that trigger cooperation~\cite{nowak04,axelrod84} constitute some of the major goals in EGT. 
In particular, the emergence of cooperation becomes really interesting when agent interactions are based on games having a Nash equilibrium of defection.
In this context, the Public Goods Game (PGG hereinafter) is one of the most famous models~\cite{perc03}, and it is based on a simple dynamics: agents play as cooperators or as defectors. Cooperators contribute to the public good providing a coin (or token), while defectors contribute nothing. Now, the set of the collected coins is enhanced by a numerical parameter called 'synergy factor', and eventually it is equally divided among all agents (no matter their strategy). Defection is the Nash equilibrium of this game, therefore it can be adopted for studying new mechanisms that support cooperation.
Here a coin, constituting the contribution provided by each cooperator in the PGG, represents a very general form of resource as money in an economical system, food in an ecological one, and so on and so forth. Thus, we assume that a contribution can be viewed as a form of free energy, having a relation with the fitness of the individual (see also~\cite{javarone03}).
In addition, a relevant aspect of evolutionary games is given by the possibility that agents can change their strategy over time. Notably, EGT is a framework for representing systems with competing strategies and, usually, rational agents tend to imitate richer or stronger opponents in order to increase the probability to gain higher payoffs at the next time steps.
In this work we focus on host-pathogen interactions~\cite{host_pathogen01}, like those we can observe in some multicellular organisms. 
The underlying complexity of these processes emerges from the interactions between bacteria and the immune system of a host organism, being the latter the result of an orchestration among several entities.
Nowadays, mathematical biology~\cite{murray01} represents one of the main attempts for describing similar scenarios both in quantitative and in analytical terms. Although a mathematical approach requires a high level of abstraction from the real scenario, a wide list of results, for instance in computational epidemiology~\cite{vespignani01,moreno02} and in genomics~\cite{genomics01}, suggests that the research in this area can be strongly helpful and promising.
Let us now focus the attention on host-pathogen interactions~\cite{moreno03}. In general, bacteria may behave as saprophyte, as commensal, or as parasite~\cite{bacteria01}.
In the proposed model, we refer to the third case (i.e. parasite).
In doing so, the host-pathogen interaction can be viewed as a two species challenge, i.e. bacteria versus cells of a host organism. Therefore, we consider this competitive biological process in a heterogeneous system.
Under this perspective, there are two kinds of equilibria: the success of one species, or a co-existence between them (we take here the opportunity for highlighting a difference between the proposed game and classical games in EGT where the actual conflict is among strategies). 
In principle, the optimal equilibrium for pathogens is represented by the co-existence, since if they completely succeed, the living organism quickly becomes inhospitable (i.e. dying).
For instance, in biological terms a widespread invasion may lead the host organism to a severe pathological state defined `blood poisoning'. 
On the other hand, a steady-state can be reached by bacteria behaving like `commensal' and `opportunistic'~\cite{bacteria02}, since they avoid a massive spread in the host organism.
Here, using the language of statistical physics~\cite{huang01}, the processes leading to the mentioned equilibria can be interpreted as order-disorder phase transitions, where the disordered phases correspond to the various forms of co-existence, while the ordered ones to the complete prevalence of one species. Therefore, even if only from a theoretical point of view, the studying of order-disorder phase transitions in these heterogeneous systems may, in principle, allow to get further insights on the considered biological phenomenon. Moreover, as shown in~\cite{sole01,sole02}, theoretical approaches to biology based on physics may constitute a fundamental ingredient for achieving a deep knowledge on the dynamics of many complex biological processes.
In this scenario, the orchestrating dynamics of the immune system and the invasive strategy of parasites, inspired us to represent their interactions as an evolutionary game between cooperators and defectors.
Thus, we introduce the Host-Pathogen game (hereinafter HPG) for modeling this complex biological challenge.
In the proposed game, the host organism is mapped to cooperative cells, while the pathogens to a population of defectors. Cells cooperate only among cells, while try to kill bacteria that, in turn, behave as defectors (even among them).
It is important to point out that agents never change strategy (here a second difference from classical games in EGT). However, as better explained later, some cells appear to behave as defectors due to their commitment in removing close bacteria.
Moreover, the spreading of pathogens comes as result of their willingness to enforce their individual wealth (i.e. their payoff).
The main dynamics of HPG entail that cells can use a fraction of their energy, corresponding to the contribution they provide to the organism, to remove close pathogens, i.e. those in their neighborhood. In turn, when a pathogen survives, it absorbs the received energy, reinforcing itself.
Let us now to observe that surviving an attack and even taking profit from it may, in principle, be viewed as the realization of a famous Nietzsche's aphorism `'That which does not kill us makes us stronger''~\cite{nietzsche01}. Broadly speaking, the latter can be also interpreted also as a principle of adaptation in a hostile environment.
Finally, as in many living systems, the temperature plays a relevant role in the HPG. For instance, generally speaking, the higher the temperature the higher the efficiency of an immune system in fighting pathogens. So, accordingly, in the proposed model high temperatures support the host organism.
The dynamics of HPG are studied by means of numerical simulations, implemented for analyzing the equilibria that can be reached, and to evaluate whether they can be related to some real scenarios (although at a very abstract level).
Before moving to the full description of this investigation, it is worth to mention some previous attempts in using evolutionary and game theory models for studying host-pathogen interactions. For instance, in~\cite{taylor01} an analytical model studies these biological interactions highlighting the relevant role of 'plasticity', in~\cite{hummert01} authors provided a nice review on evolutionary games where cells are mapped to players, and authors of~\cite{tago01} presented an interesting model based on game theory for understanding these complex biological dynamics in terms of Nash equilibrium.
The remainder of the paper is organized as follows: Section~\ref{sec:model} introduces the proposed model. Section~\ref{sec:results} shows results of numerical simulations. Eventually, Section~\ref{sec:conclusions} ends the paper.
\section{The Host-Pathogen Game}\label{sec:model}
\begin{figure}[h!]
\centering
\includegraphics[width=0.8\textwidth]{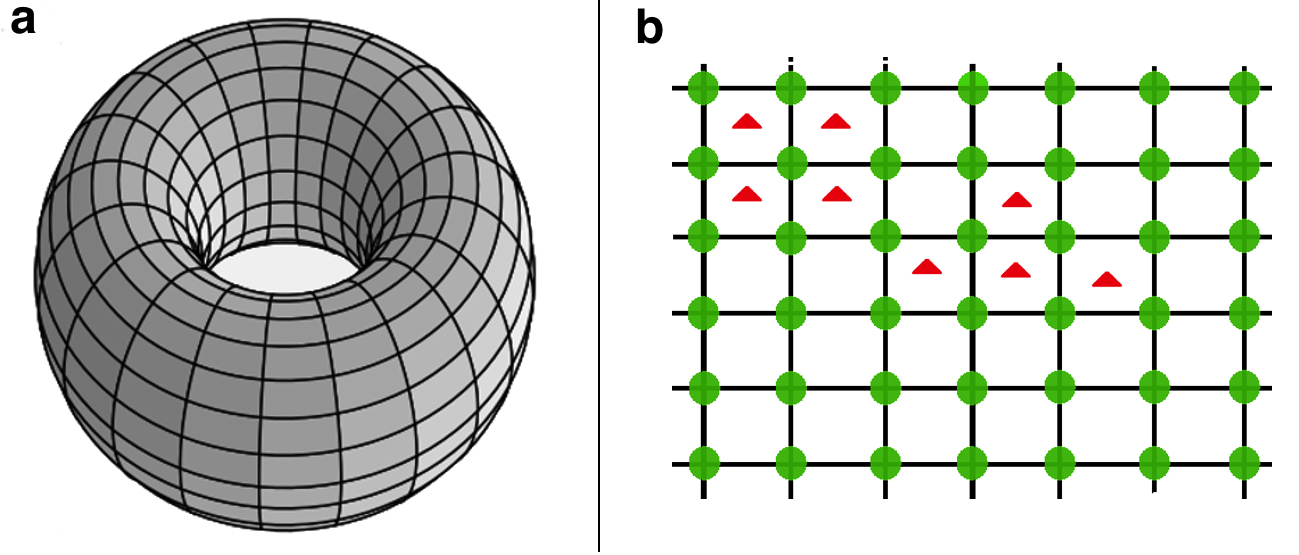}
\caption{\small Pictorial representation of the interaction topology. In particular, cooperative cells are arranged over a toroid (i.e. \textbf{a}), while bacteria occupy inner squares. Notably, \textbf{b} shows a fragment of the toroid, with green circles representing cooperative cells, and red triangles representing bacteria inside infected squares. \label{fig:topology}}
\end{figure}

The HPG aims to model the competitive dynamics of a spatially structured~\cite{moreno01,pena01} two-species population, i.e. cells of a host organism versus external agents (e.g. bacteria or parasites).
The former are arranged in a square lattice with continuous boundary conditions, while the latter occupy the inner squares of the lattice ---see Figure~\ref{fig:topology}. In doing so, when a bacterium invades a square, the four cells (located at the vertices of the infected square) try to remove it.
Cells behave as cooperators, whereas bacteria as defectors.
Therefore, as in the PGG, cooperators provide a unitary amount of a resource. In this specific case, the resource corresponds to a form of free energy (due to the high level of abstractness of the proposed model hereinafter we take the freedom to use just the term 'energy'), used by the host organism to carry out its living functions, or other tasks. Then, cells indirectly receive a payoff (e.g. oxygen) when this biological engine correctly works.
Now, once a square gets infected, cells at its vertices use a fraction of the energy contribution for trying to remove the external agents.
Since a cell is surrounded by four squares (see Figure~\ref{fig:topology} where cells are represented by green circles, placed in the vertices of the squares), it may reserve only $\frac{1}{4}$ of its energy contribution for each square.
As result the contribution of a cell, to the host organism, decreases as the number of nearest infected squares increases. Thus, when cells try to remove one or more parasites, indirectly behave as defectors. This observation recovers a particular importance since the adopted dynamics let emerge defectors even in the cell population. Remarkably, defection among cells comes as an indirect effect, because they always provide the same amount of energy but, in this particular case (i.e. defection), it is targeted towards the external agent.
Due to its relevance, this phenomenon requires a further description. When a cell has only one nearest infected square, its energy contribution is scarcely reduced (i.e. by $\frac{1}{4}$); on the other hand, when a cell is completely surrounded by infected squares, its contribution is totally used for fighting parasites (i.e. a quarter for each infected square).
In turn, since each square has four vertices (one per cell), a parasite is always attacked by means of a unitary amount of energy, coming from the summation of the fractions of energy provided by the four cells, each equal to $\frac{1}{4}$ (whose summation gives $1$).
Now, using a Glauber-like approach, we define the probability $p^h$ to reduce the payoff of a parasite (until its removal from a square), and that to fail (i.e. $p^b$) as follows
\begin{equation}\label{eq:probabilities}
\begin{cases}
p^{h} = e^{-\beta H}\\
p^{b} = 1 - p^{h}
\end{cases}
\end{equation}
\noindent with $H = \sum_{i=1}^{4} \frac{c^i}{4} = 1.0$, where $c$ is the energy contribution of a single cell, and $\beta = \frac{1}{(T - T_c)\cdot \alpha}$. 
Here, $H$ can be viewed as the Hamiltonian representing the energy transferred to a square and with constant unitary value (i.e. $1$), $\alpha$ as the Boltzmann constant (set to $0.7$), $T_c$ as the minimum temperature a living organism may have to work physiologically, and $T$ the actual temperature of an organism.
The equations~\ref{eq:probabilities} highlight the significant role of the temperature in the HPG, as for high $T$ the $p^{h} \to 1$, while for low $T$ the opposite occurs (i.e. parasites are never removed).
Furthermore, as briefly mentioned before, once a parasite survives an attack it can use the absorbed energy for increasing its payoff.
In particular, when a parasite infects a square, its payoff is still equal to zero. Then, in order to increase its payoff, the parasite has to infect all the nearest squares (i.e. the adjacent ones). Thus as an attack of the host cells fails, the parasite spreads if there is a nearest square free (i.e. not infected), otherwise it increases its own payoff. Therefore, the spreading dynamics is limited by the spatial constraints of the considered environment (i.e. a toroid) since, when all squares get infected, parasites may only increase their payoff. At the first glance, this assumption can appear far from a real biological scenario. However, it is worth to note that the increasing of the payoff of the parasites can be viewed as the increasing of their resistance towards external attacks.
Thus the payoff has a special role, and a bacterium (or parasite) can accumulate it over time behaving as a memory-aware agent (see~\cite{javarone02}), reducing the probability to be removed from the host organism during new attacks. Here, we observe a further element of contact between the proposed model and real scenarios, i.e. when bacteria survive to antibiotic treatments, they become more resistant to that drug (i.e. the administered antibiotic).
Instead, every time a cell succeeds according to~\ref{eq:probabilities}, the payoff of the attacked parasite reduces of $1$, so that it is removed only once its payoff becomes negative.
The possibility to conserve the payoff over time entails the parasite is able to enforce itself (e.g. creating a kind of shield). As previously reported, this phenomenon can be viewed in terms of a form of adaptation to a hostile environment, or as the formation of a 'callus' resulting from a continuous mechanical stimulus in the skin.
Eventually, it is worth to emphasize the role of the temperature in HPG. As previously described, the increasing of the temperature results from a neural mechanism whose aim is to trigger the immune-system. In the proposed model, the neural mechanism has not been considered, while the temperature is included in the parameter $\beta$ of Eq.~\ref{eq:probabilities}. So, tuning its value, one can analyze the outcomes of the game according to the reached equilibrium (i.e. the increasing or decreasing of parasites). Then, in HPG, increasing the temperature allows to increase the probability to remove the parasites, so that temperature supports cooperators (cells).
Summarizing, the HPG can be described as follows:
\begin{enumerate}
\item At $t=0$ a number of squares get infected by parasites (with a payoff equal to zero);
\item While the number of infected squares is greater than $0$:
\item \_\_ Randomly select a parasite (i.e. focusing on a specific infected square);
\item \_\_ The parasite reduces its payoff by $1$ with probability $p^h$. So, if its payoff becomes negative the parasite is removed and the square gets healthy, otherwise: 
\item \_\_\_\_ IF all nearest squares are infected: the parasite increases by $1$ its payoff;
\item \_\_\_\_ ELSE: randomly select a free (i.e. healthy) nearest square and infect it with a new parasite (whose payoff is set to $0$);
\item repeat from $(2)$ until the population reaches an ordered phase, or up to a (limited) number of time steps elapsed.
\end{enumerate}
Before to show results of numerical simulations, we observe that if agents were not 'memory-aware' (i.e. not able to save their payoffs), the amount of infected squares $s$ could be analytically computed according to the following equation
\begin{equation}\label{eq:analytical}
\frac{ds}{dt} = s\cdot(p^b - p^h)
\end{equation}
Whose solution reads
\begin{equation}\label{eq:analytical_solution}
s(t) = s(0)\cdot e^{z t}
\end{equation}
\noindent with $z = (p^b - p^h)$ and $s(0)$ is the initial amount of infected cells in the organism. Here, in principle, some conditions could lead the system towards an oscillatory dynamics like the one emerging in models like that of Lotka-Volterra. However, as before described, the complexity of the HPG entails numerical simulations are mandatory to investigate its behavior.
\section{Results}\label{sec:results}
Numerical simulations have been performed for analyzing two different cases: constant temperature and varying temperature, always considering a basic temperature equal to $T = 33$ (previously indicated as $T_c$). The first case allows to study the evolution of the heterogeneous population (i.e. cells and parasites) without to vary the system temperature (i.e. that of the whole host organism). The second case allows to analyze the role of a variant temperature in this dynamics (e.g. as a reaction of the organism to the infection). Eventually we recall that, in both cases, cells are arranged in a bi-dimensional square lattice, with continuous boundary conditions, and parasites are arranged inside the inner squares.
\subsection*{Constant Temperature}
\begin{figure}[h!]
\centering
\includegraphics[width=0.8\textwidth]{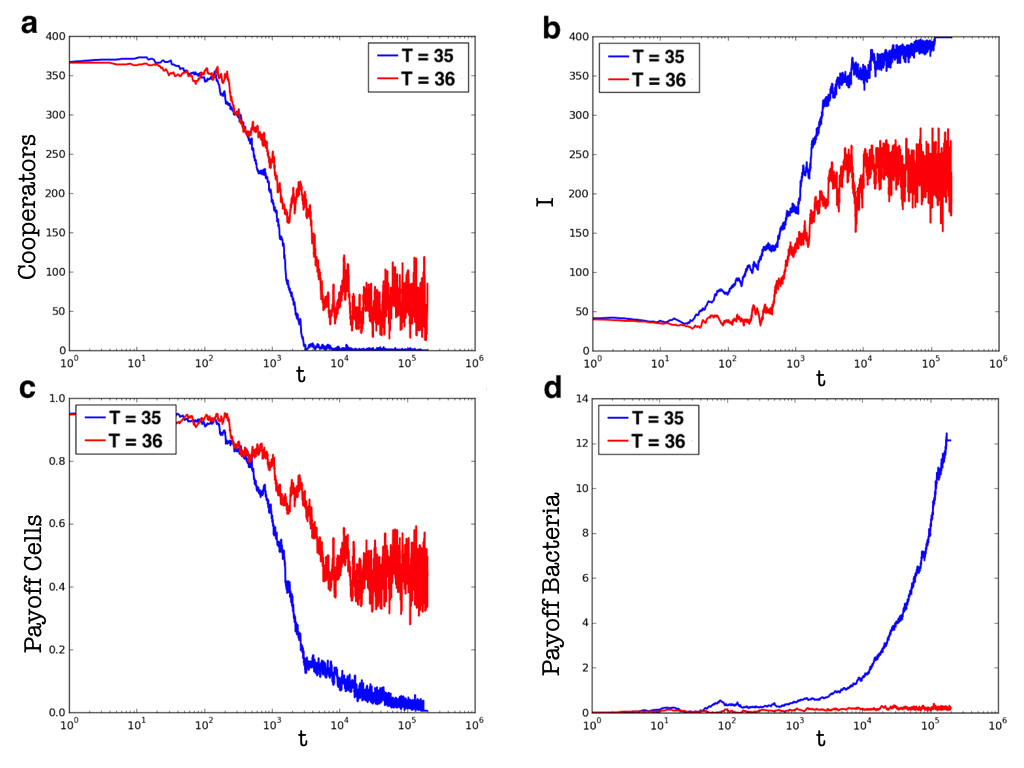}
\caption{\small Plots \textbf{a} and \textbf{b} represent the number of cooperators and that of infected cells ($I$), respectively, over time. Plots \textbf{c} and \textbf{d} represent the payoff gained by cells and by parasites over time, respectively. In all plots, the blue line represents results achieved by $T = 35$, while the red one those achieved by $T = 36$. Results are averaged over different runs. \label{fig:figure_3}}
\end{figure}

Let us consider the first case, i.e. the evolution of the population at constant temperature. Due to the computational effort required to simulate the proposed model, we consider a population of $N = 400$ cells, with an initial density of bacteria corresponding to the $5\%$ of cells. Two different temperatures are compared: $T = 35$ and $T = 36$. Here, we emphasize that, although we studied the proposed model while considering also different initial configurations (e.g. different temperatures and initial densities of bacteria), those reported are the most significant.
\begin{figure}[h!]
\centering
\includegraphics[width=0.8\textwidth]{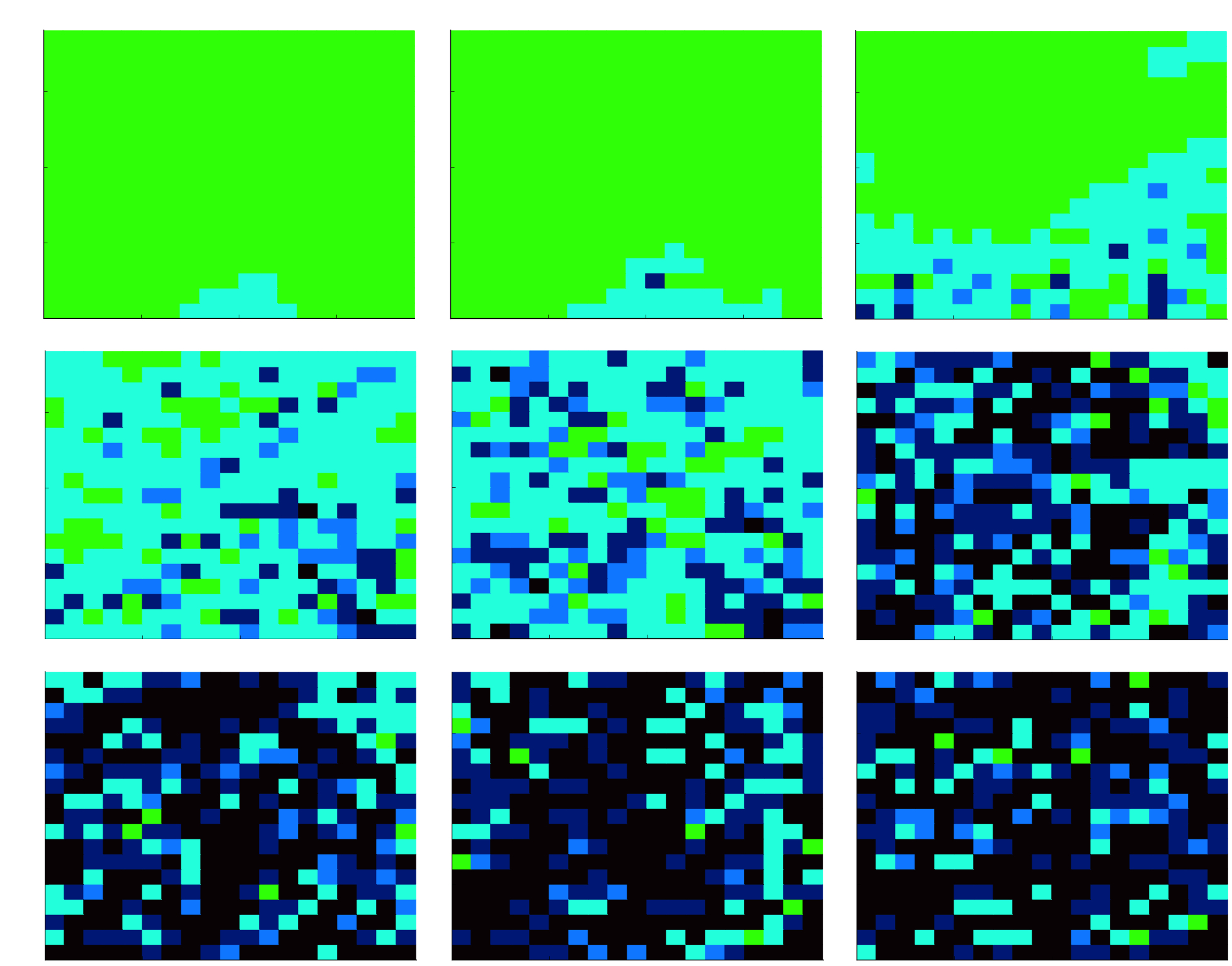}
\caption{\small Evolution of the agent population, composed of $N = 400$ cells, with a constant temperature $T = 35$. At the beginning, the amount of parasites corresponds to the $5\%$ of cells. Colors represent the state of cells: those green are healthy (i.e. not infected by parasites), while those cyan, blue and black are infected. The darkness of infected cells indicates the parasites' payoff, i.e. the darkest the richest (i.e. the strongest versus cell attacks). Each subplot refers to a different time step, starting from $t  = 0$ (i.e. the first one) up to $t = 175000$ (from left to right, and from the upper part to the lower part). All plots refer to a single realization of a simulation. \label{fig:figure_1}}
\end{figure}
\begin{figure}[h!]
\centering
\includegraphics[width=0.8\textwidth]{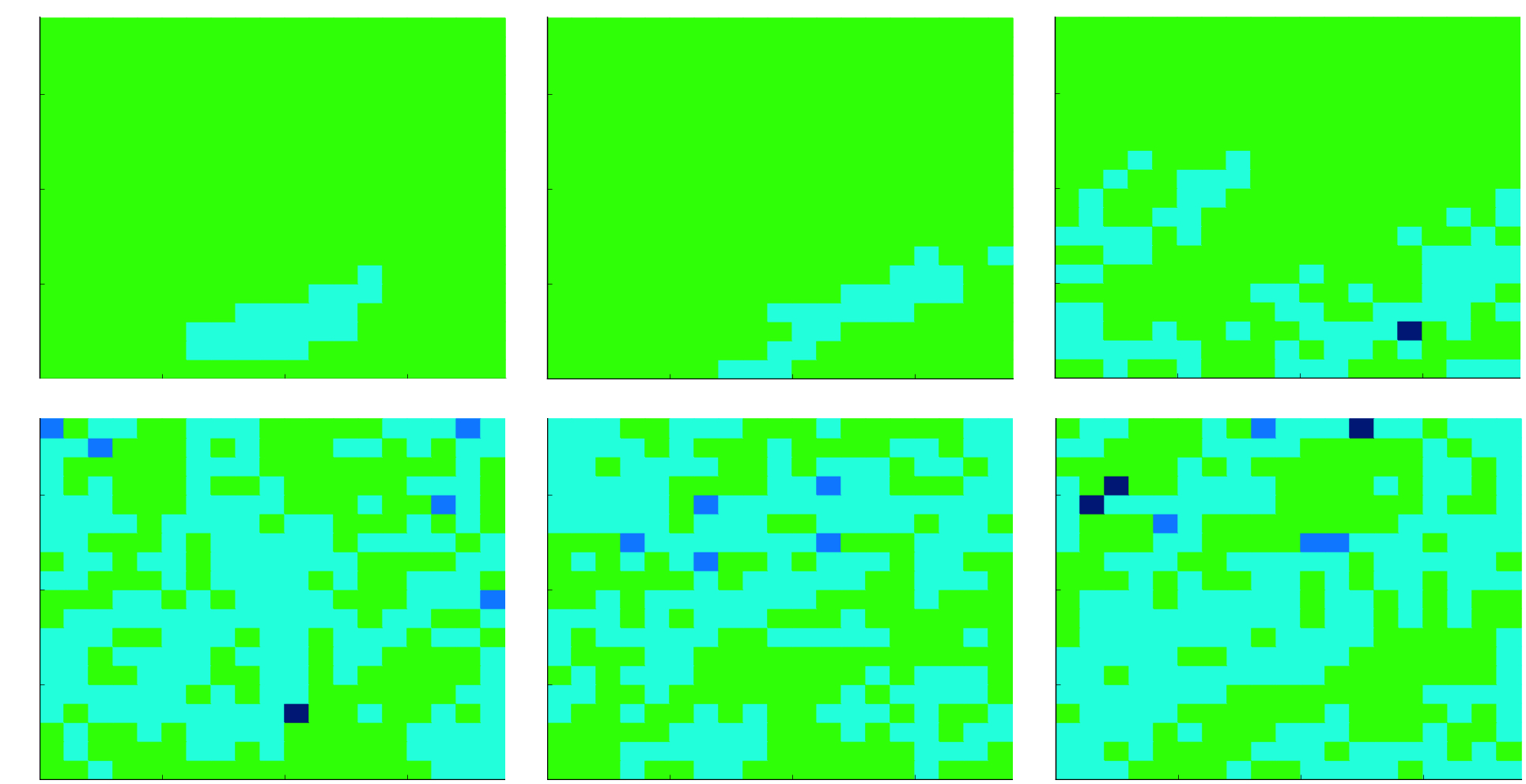}
\caption{\small Evolution of the agent population, composed of $N = 400$ cells, with a constant temperature $T = 36$. At the beginning, the amount of parasites corresponds to the $5\%$ of cells. Colors represent the state of cells: those green are healthy (i.e. not infected by bacteria), while those cyan, blue and black are infected. The darkness of infected cells indicates the parasites' payoff, i.e., the darkest the richest (i.e. the strongest versus cell attacks). Each subplot refers to a different time step, starting from $t  = 0$ (i.e. the first one) up to $t = 200000$ (from left to right, and from the upper part to the lower part). All plots refer to a single realization of a simulation. \label{fig:figure_2}}
\end{figure}
Figures~\ref{fig:figure_1} and~\ref{fig:figure_2} show results related to the temperatures $T = 35$ and $T = 36$, respectively. Even if, in both cases, parasites are able to spread in the whole organism, the first one (i.e. Figure~\ref{fig:figure_1}) indicates that at $T = 35$ parasites completely prevail, while the second (i.e. Figure~\ref{fig:figure_2}) shows the emergence of a steady-state characterized by the co-existence between the two species.
Figure~\ref{fig:figure_3} reports the density of cooperators and defectors in the cell population, over time (plots \textbf{a} and \textbf{b}), and the payoff gained by cells (plot \textbf{c}) and by parasites (plot \textbf{d}).
In all figures, we compare results achieved by setting the organism temperature to $T= 35$ (blue line) and to $T= 36$ (red line). As we can observe, Figure~\ref{fig:figure_3} confirms qualitative results shown in Figure~\ref{fig:figure_1} and Figure~\ref{fig:figure_2}, i.e. increasing the organism temperature, by only one degree, the scenario radically changes. In particular, at $T = 35$, parasites prevail completely causing what that can be interpreted as a 'blood poisoning' in living systems, while at $T= 36$ a steady-state of coexistence is reached, even if parasites spread in many cells. 
\subsection*{Varying Temperature}
\begin{figure}[h!]
\centering
\includegraphics[width=0.8\textwidth]{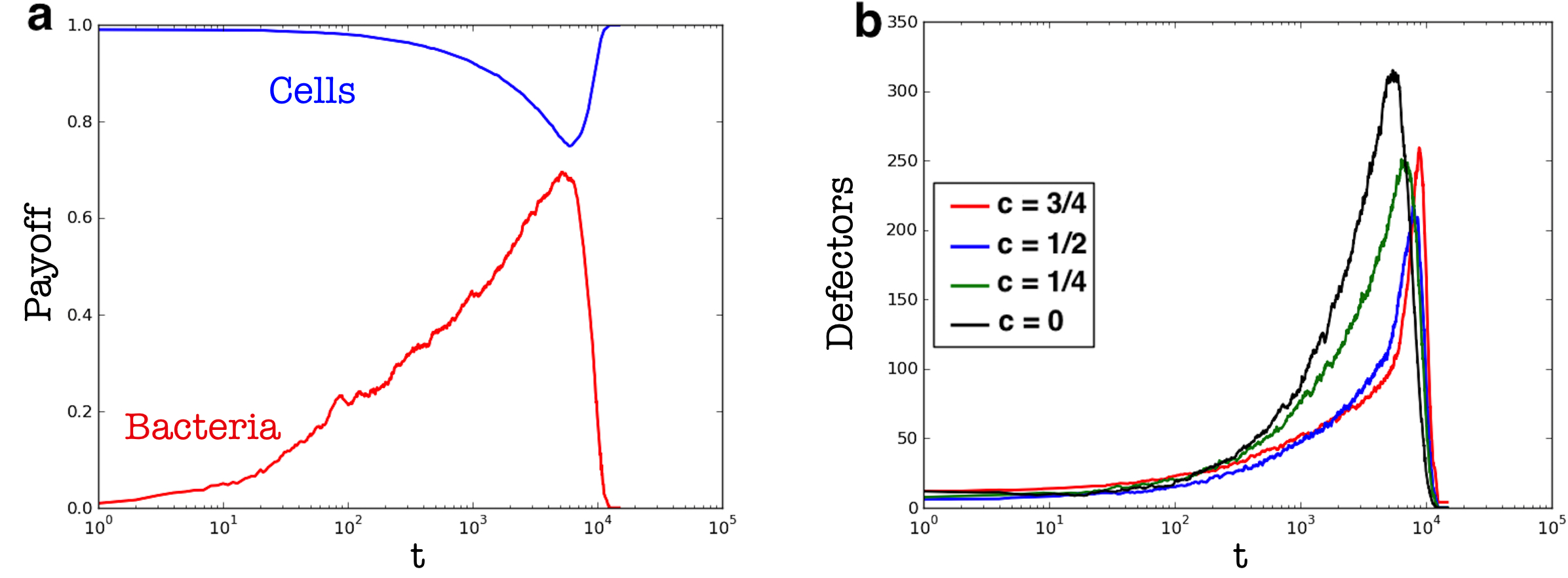}
\caption{\small Variation of the payoff over time. \textbf{a} Payoff of cells (blue line) versus that of parasites (red line). \textbf{b} Fraction of defectors among cells. Cells are induced to defect by reducing their energy support to the organism in different fractions. Red, blue and green lines indicates partial cooperators providing, respectively, a contribution to the organism equal to: $\frac{3}{4}$, $\frac{1}{2}$, $\frac{1}{4}$. Instead, black line indicates full defectors, i.e. cells that use the whole energy contribution to fight parasites. Results are averaged over different runs. \label{fig:figure_5}}
\end{figure}

Now we show results achieved by considering a varying temperature of the host organism. This process is of interest since it occurs in many biological systems, i.e. once an organism is infected by a parasite or by an external agent (being living or not) the host increases the temperature in order to improve its efficiency in restoring the healthy condition. However, it is worth to mention that the increasing of the host temperature can be caused by different factors and, in general, is the outcome of a complex mechanism.
In this case, we considered a population of $N = 2500$ cells, starting with two different initial temperatures: $T = 35$ and $T = 36$. The introduction of a varying temperature reduces the computational effort to simulate the model so that, as one may observe, the current population has more cells than that implemented in the previous case. Finally, as before, the basic temperature is $T_c = 33$ (i.e. the minimum physiological temperature in centigrade degrees for having a working organism, according to our very general assumptions on living systems).
Now the starting density of parasites is equal to the $1\%$ of cells, but the temperature can increase only once that the density of parasites is equal (or greater) than $25\%$ of cells. The heating is slow, i.e. the temperature increases of a $\Delta T = 0.0001$ at each time step once the density condition, before described, is reached. 
\begin{figure}[h!]
\centering
\includegraphics[width=0.8\textwidth]{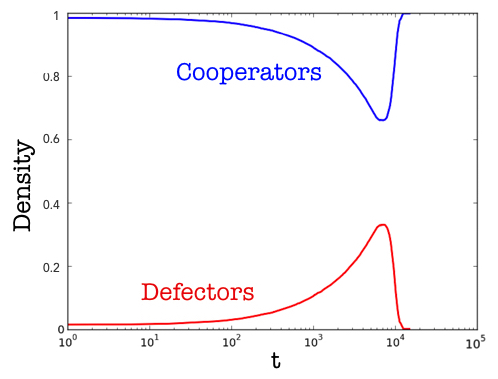}
\caption{\small Density of cooperators (blue) and of defectors (red) over time. Results are averaged over different runs. \label{fig:figure_4_a}}
\end{figure}
\begin{figure}[h!]
\centering
\includegraphics[width=0.8\textwidth]{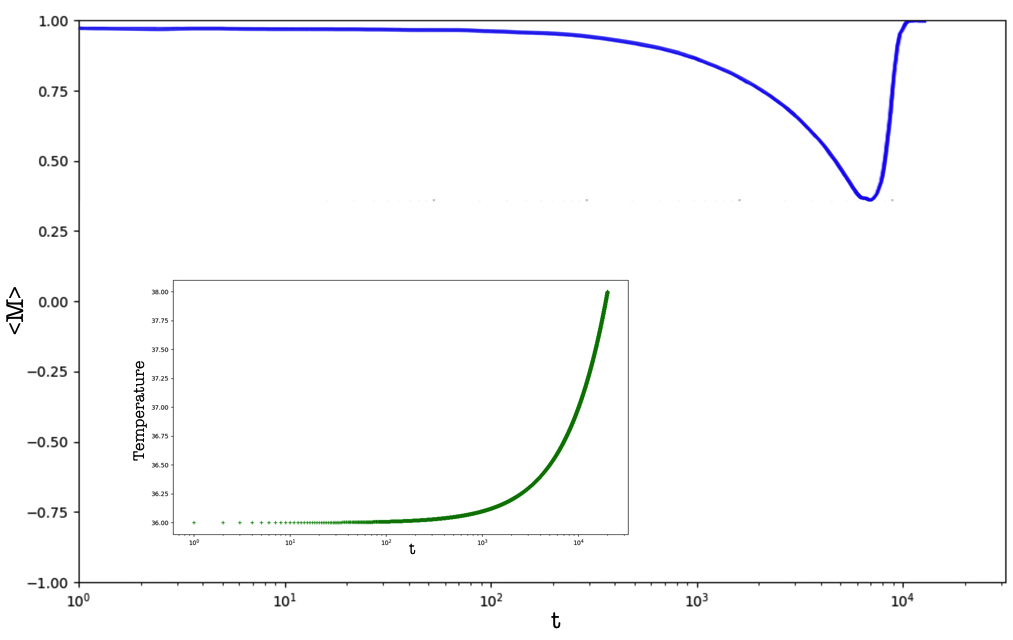}
\caption{\small Average magnetization of the population over time. The inset shows the heating processes applied to the system. Results are averaged over different runs.  \label{fig:figure_4_b}}
\end{figure}
Figure~\ref{fig:figure_4_a} shows the density of cooperators and of defectors over time, and Figure~\ref{fig:figure_4_b} the related average magnetization (the inset reports the variation of the Temperature). The latter is computed considering a binary variable $\sigma = \pm 1$, being $+1$ for cooperative cells, and $-1$ for defector cells. 
In so doing, the magnetization (see~\cite{mobilia01}) reads
\begin{equation}\label{eq:magnetization}
M=\frac{1}{N}\sum_{i=1}^N \sigma_i
\end{equation}
We remind that parasites are always defectors, so they have not been considered in the computation of the average magnetization (equation~\ref{eq:magnetization}).
From the point of view of evolutionary game theory, it is possible to observe an interesting phenomenon in the dynamics of the proposed model: once that cells are surrounded by parasites, their energy contribution is used to fight them; therefore these cells behave like defectors, since cannot provide a full contribution to the organism. In few words, the spreading of parasites indirectly lets emerge defectors among cells that, involuntarily, are no more perceived as cooperators.
In Figure~\ref{fig:figure_5} we can observe this phenomenon with more details.
In particular, the plot \textbf{b} (of Figure~\ref{fig:figure_5}) shows the emergence of different kinds of defectors: from those that are 'partial defectors' to those that are 'full defectors'. The difference between 'partial' and 'full' defector is related to the amount of energy these cells are able to provide to the organism. Notably, a full defector is a defector as meant in EGT, while a partial defector corresponds to an agent that provides only a fraction of the energy contribution.
\section{Discussion and Conclusion}\label{sec:conclusions}
In this work we study the dynamics of complex biological processes, i.e. host-pathogen competitions, by the framework of Evolutionary Game Theory.
In particular, we consider a two-species competition, i.e. cells of a host organism versus bacteria (or parasites).
The proposed model, named Host-Pathogen Game, strongly simplifies real biological scenarios. However, the related dynamics, investigated by means of numerical simulations, show that different equilibria, similar on a quality level to those observable in real contexts, can be reached.
For instance, both the complete removal of parasites, and the invasive spreading of these external agents, correspond to healing processes and deep infections mechanisms (like 'blood poisoning'), respectively.
Even if agents of HPG never change strategy, the interactions between cells and parasites can lead the former to behave actually as defectors. So, we observe an evolutionary mechanism introduced by the invasion and the spreading of parasites in the host organism, that entails the emergence of (false) defectors among cells.  
In our view, the relevance of this observation lies in the fact that HPG introduces the possibility to become defectors while behaving as cooperators (i.e. while still paying a contribution). In particular, the strategy of an agent is not the one adopted, but that perceived by its neighbors. Therefore, we deem that this phenomenon might deserve further investigations also beyond its biological interpretation.
Eventually, as in many living organisms, also in HPG the system temperature plays a key role: high values allow cells to prevail, while low ones support bacteria. Thus, this point represents a further link between the proposed model and biological scenarios.
Here, it is useful to emphasize that, while in the proposed model we assume that the temperature cannot be negative, for some particular organisms nature shows a richer scenario (see~\cite{clarcke01} for further details). For that reason, we consider only a limited range of temperatures that, in the centigrade scale, are considered physiological for many animals, e.g. several mammalians. In addition, in the second case, we focused only on slow variations of the temperature. However, the rate of variation of the latter can be a further interesting parameter to consider in future analyses (e.g. for evaluating the effects of a thermal shock).
To conclude, the adopted simplifications for modeling a complex process are rewarded by results that can have a biological interpretation (e.g. 'blood poisoning'). However, considering the assumptions and the level of abstraction adopted in the model, we think that the preliminary results achieved in this work may be suitable for describing biological competitions involving very simple multicellular organisms versus bacteria. Therefore, further developments are required for describing the dynamics within more complex systems like vertebrates.   
Moreover, a further development of the proposed model could lie in the introduction of some form of cooperation among parasites, as can actually be observed in some real cases. 
In the light of these considerations, we deem relevant to point out that Evolutionary Game Theory may actually constitute a promising framework for analyzing also complex biological phenomena, and we suggest that further studies would be important for shedding new light on different aspects of the proposed model.
\section*{Conflict of Interest Statement}
The author is employed by company nChain LTD, and he declares that the research was conducted in the absence of any commercial or financial relationships that could be construed as a potential conflict of interest.

\section*{Acknowledgments}
The author developed the first part of this work at the University of Cagliari (Dept. of Mathematics and Computer Science) Italy, and at the University of Hertfordshire (School of Computer Science) UK. In addition, the author has been partially supported by the National Group of Mathematical Physics (GNFM-INdAM).


\begin{thebibliography}{99}

\bibitem{nowak01}
Nowak, M.A. and May, R.M.:
Evolutionary games and spatial chaos.
\emph{Nature} \textbf{359} 826--829 (1992)

\bibitem{perc01} 
Perc, M., Grigolini, P.:
Collective behavior and evolutionary games – An introduction.
\emph{Chaos, Solitons \& Fractals}  \textbf{56} 1–5 (2013)

\bibitem{perc02} 
Szolnoki, A., Perc, M.:
Collective influence in evolutionary social dilemmas.
\emph{EPL} \textbf{113-5} 58004 (2016) 

\bibitem{szolnoki01} 
Szolnoki, A., Perc, M.:
Conformity enhances network reciprocity in evolutionary social dilemmas.
\emph{J. R. Soc. Interface} \textbf{12} 20141299 (2015) 

\bibitem{masuda01} 
Tanabe, S., Suzuki, H., Masuda, N.:
Indirect reciprocity with trinary reputations.
\emph{Journal of theoretical biology} \textbf{317} 338--347 (2013)

\bibitem{tomassini01} 
Tomassini, M.:
Introduction to evolutionary game theory.
\emph{Proc. Conf. on Genetic and evolutionary computation companion} (2014)

\bibitem{moreno01} 
Perc, M., Gomez-Gardenes, J., Szolnoki, A., Floria, L.M., and Moreno, Y.:
Evolutionary dynamics of group interactions on structured populations: a review.
\emph{J. R. Soc. Interface} \textbf{10-80} 20120997 (2013)

\bibitem{traulsen01}
Traulsen, A., Reed, F.A:
From genes to games: cooperation and cyclic dominance in meiotic drive.
\emph{Journal of theoretical biology} \textbf{299} 120-125 (2012)

\bibitem{szabo01} 
Hauert, c. and Szabo, G.:
Game theory and Physics.
\emph{A. J. Phys.} \textbf{73 - 405} (2005)

\bibitem{friedman01}
Friedman, D.:
On economic applications of evolutionary game theory.
\emph{Journal of Evolutionary Economics} \textbf{8-1} 15--43 (1998)

\bibitem{nowak03}
Fu, F., Rosenbloom, D.I. Wang, L., Nowak, M.A.:
Imitation dynamics of vaccination behaviour on social networks.
\emph{Proc. R. Soc. B} \textbf{278} 42--49 (2011)

\bibitem{shuster01}
Schuster, S., de Figueiredo, L., Schroeter, A., and Kaleta, C.:
Combining metabolic pathway analysis with evolutionary game theory. Explaining the occurrence of low-yield pathways by an analytic optimization approach.
\emph{BioSystems} \textbf{105} 147--153 (2011)

\bibitem{frey01}
Frey, E.:
Evolutionary game theory: theoretical concepts and applications to microbial communities. 
\emph{Physica A} \textbf{389} 4265–-4298 (2010)

\bibitem{javarone01} 
Javarone, M.A., Antonioni, A., Caravelli, F.:
Conformity-Driven Agents Support Ordered Phases in the Spatial Public Goods Game.
\emph{EPL} \textbf{114-3} (2016)

\bibitem{szolnoki02} 
Szolnoki, A., Perc, M.:
Zealots tame oscillations in the spatial rock-paper-scissors game.
\emph{Phys Rev E} \textbf{93} 062307 (2016)

\bibitem{javarone04} 
Javarone, M.A.:
Statistical Physics and Computational Methods for Evolutionary Game Theory
\emph{Springer} \textbf{10.1007/978-3-319-70205-6} (2018)

\bibitem{nowak04}
Nowak, M.A.:
Five rules for the evolution of cooperation.
\emph{Science} \textbf{314-5805} 1560--1563 (2006)

\bibitem{axelrod84}
Axelrod, R.:
The Evolution of Cooperation.
\emph{Basic Books} (1984)

\bibitem{perc03} 
Szolnoki, and Perc, M.:
Reward and cooperation in the spatial Public Goods Game.
\emph{EPL} \textbf{92} 38003 (2010)

\bibitem{javarone03} 
Javarone, M.A.:
Solving Optimization Problems by the Public Goods Game.
\emph{EPJ-B} \textbf{90-171} (2017)

\bibitem{host_pathogen01} 
Rescigno, M., and Borrow, P.:
The Host-Pathogen Interaction. New Themes from Dendritic Cell Biology
\emph{Cell} \textbf{106-3} 267–-270 (2001)

\bibitem{murray01}
Murray JD:
Mathematical Biology.
\emph{Springer} (2002)

\bibitem{vespignani01} 
Pastor-Satorras, R., Vespignani, A.:
Epidemic spreading in scale-free networks.
\emph{PRL} \textit{86-14} 3200 (2001)

\bibitem{moreno02} 
Moreno, Y., Pastor-Satorras, R., Vespignani, A.:
Epidemic outbreaks in complex heterogeneous networks.
\emph{EPJ-B} \textit{26-4} 521-529 (2002)

\bibitem{genomics01} 
Allendorf, F.W., Hohenlohe, P.A., Luikart, G.:
Genomics and the future of conservation genetics.
\emph{Nature Reviews Genetics} \textit{11} 697-709 (2010)

\bibitem{moreno03} 
Poletto, C., Meloni, S., Colizza, V., Moreno, Y., Vespignani, A.:
Host Mobility Drives Pathogen Competition in Spatially Structured Populations.
\emph{PLoS Comput Biol} e1003169 (2013)

\bibitem{bacteria01} 
Snyder, L., Peters, J.E., Henkin, T.M., Champness, W.:
Molecular Genetics of Bacteria.
\emph{ASMscience} (2013)

\bibitem{bacteria02} 
Macpherson, A.J., Harris, N.L.:
Interactions between commensal intestinal bacteria and the immune system.
\emph{Nature Reviews Immunology} \textit{4} 478-485 (2004)

\bibitem{huang01}
Huang, K.:
Statistical Mechanics.
\emph{Wiley 2nd Ed.} (1987)

\bibitem{sole01} 
Duran-Nebreda, S., Montanez, N., Bonforti, A., Sole\', R.:
The Paths to Artificial Multicellularity: From Physics to Evolution.
\emph{MIT press} Multicellularity: Origins and Evolution (2016)

\bibitem{sole02} 
Sole\', R.:
Synthetic Transitions: Towards a New Synthesis.
\emph{SFI WORKING PAPER} (2016)

\bibitem{nietzsche01} 
Friedrich Wilhelm Nietzsche:
Twilight of the the Idols
(1876)

\bibitem{taylor01} 
Taylor, P.D. et al.:
The evolutionary consequences of plasticity in host–pathogen interactions.
\emph{Theoretical Population Biology} \textbf{69}  323--331 (2006)

\bibitem{hummert01} 
Hummert, S. et al.:
Evolutionary game theory: cells as players.
\emph{Molecular BioSystems} \textbf{10}  3044--3065 (2014)

\bibitem{tago01} 
Tago, D., Meyer, D.F.:
Economic Game Theory to Model the Attenuation of Virulence of an Obligate Intracellular Bacterium.
\emph{Frontiers in Cellular and Infection Microbiology} \textbf{6}  86 (2016)

\bibitem{pena01} 
Pena, J., Noldeke, G., Lehmann, L.:
Evolutionary dynamics of collective action in spatially structured populations.
\emph{Journal of Theoretical Biology} \textit{382} (2015)

\bibitem{javarone02} 
Javarone, M.A.:
Statistical Physics of the Spatial Prisoner's Dilemma with Memory-Aware Agents.
\emph{EPJ-B} \textit{89-42} (2016)

\bibitem{mobilia01}
Mobilia, M. and Redner, S.:
Majority versus minority dynamics: Phase transition in an interacting two-state spin system.
\emph{Phys. Rev. E} \textbf{68-4} 046106 (2003)

\bibitem{clarcke01} 
Clarke, A., Morris, G.J., Fonseca, F., Murray, B.J., Acton, E., Price, H.C.:
A Low Temperature Limit for Life on Earth.
\emph{PLoS ONE} \textbf{8:6} e66207 (2013)

\end{thebibliography}
\end{document}